\documentclass[pra,twocolumn,tightenlines,showpacs,nofootinbib]{revtex4}
\usepackage{bm,dcolumn,amsmath,graphicx,amsfonts,amssymb}

\newcommand{\eref}[1]{Eq.~(\ref{#1})}
\newcommand{\tref}[1]{Table~\ref{#1}}

\begin{document}
\title{Sensitivity of the isotopologues of hydronium
to variation of the electron-to-proton mass ratio}
\author{M. G. Kozlov$^1$}
\author{S. G. Porsev$^{1,2}$}
\author{D. Reimers$^2$}
\affiliation{$^1$ Petersburg Nuclear Physics Institute, Gatchina,
Leningrad district, 188300, Russia} \affiliation{$^2$ Hamburger
Sternwarte, Universit\"{a}t Hamburg, Hamburg 21029, Germany}

\date{ \today }

\begin{abstract}
We study the sensitivity of the microwave and submillimeter transitions of the
isotopologues of hydronium to the variation of the electron-to-proton mass
ratio $\mu$. These sensitivities are enhanced for the low frequency mixed
inversion-rotational transitions. The lowest frequency transition (6.6 GHz)
takes place for isotopologue H$_2$DO$^+$ and respective sensitivity to
$\mu$-variation is close to 200. This is about two orders of magnitude larger
than the sensitivity of the inversion transition in ammonia, which is
currently used for the search of $\mu$-variation in astrophysics.
\end{abstract}
 \maketitle

\section{Introduction.}

The search for the possible variation of fundamental constants is
a poor man's way to look
for physics beyond the Standard Model. There are two main approaches
to such studies. First is to make laboratory measurements with
different atomic and molecular clocks in order to look for local
variations on the time scale of the order of a year. Second is to
compare laboratory measurements with astrophysical observations,
where one can study variations on a much bigger space and time scale
up to 10 billion years. Advantage of the first approach is in
unprecedented accuracy of experiments with atomic clocks. On the
other hand, the second approach allows exploring bigger scales and
different environments. For example, the local matter density in the
laboratory experiments and in interstellar molecular clouds differ
by more than 10 orders of magnitude. We can conclude that these two
approaches are complementary to each other \cite{FK09}. In this
paper we do not discuss laboratory experiments and focus on
astrophysical applications.

In astrophysics all frequency shifts with respect to the laboratory
values are conventionally interpreted as Doppler shifts, $\Delta
\omega/\omega = V/c$, where $V$ is the line of sight velocity of the
object and $c$ is the speed of light. However, if we observe several
lines from the same object, we can study (small) velocity offsets
between them. In the search for the variation of fundamental
constants we look for the correlation between these offsets and
sensitivities of the observed transitions to the variation to
fundamental constants.

It is now well known that the tunneling transition in ammonia is
highly sensitive to the electron-to-proton mass ratio $\mu=m_e/m_p$
\cite{VKB04}. This transition is often observed from the
interstellar dark molecular clouds in our Galaxy. Comparison of the
apparent Doppler shift for this transition with those of rotational
transitions of other molecules provides a very sensitive test for
the dependence of the mass ratio $\mu$ on the local matter density
\cite{LMK08,MLK09,LLH10}. Moreover, ammonia was detected in two high
redshift objects with $z=0.68$ and $z=0.89$. That allows to place
the most stringent present limits on $\mu$-variation on a
cosmological time scale \cite{FlaKoz07,MFMH08,HMM09,Kan11}.

Because all observed ammonia transitions have the same sensitivity
to the variation of the fundamental constants, one has to use
rotational transitions in \textit{other} molecules as a reference.
This can lead to the systematic effects caused by different spacial
distribution of the molecules in gas clouds (so called Doppler
noise) \cite{LML10}. Therefore, it is desirable to find molecules,
where one can simultaneously observe several lines with high
\textit{and} different sensitivities to $\mu$-variation. In this
case the Doppler noise is minimized because all lines are observed
from the same gas. By now we know several such molecules, including
OH \cite{CK03,Dar03} and CH \cite{Koz09}, partly deuterated ammonia
\cite{KozLapLev10}, hydronium ion (H$_3$O$^+$) \cite{KozLev11}, and
methanol (CH$_3$OH) \cite{JXK11,KL11a}. These molecules often have
higher sensitivity than ammonia and can be used as independent
source of information on variations of fundamental constants.

In this paper we consider the hydronium isotopologues H$_3$O$^+$,
H$_2$DO$^+$, HD$_2$O$^+$, and D$_3$O$^+$. Like ammonia, hydronium
also has a double minimum vibrational potential. The inversion
transitions occur when the oxygen nucleus tunnels through the plane
of the hydrogen/deuterium nuclei. This leads to an inversion
splitting of the rotational levels. The splitting in H$_3$O$^+$ is
very large, about 55 cm$^{-1}$, as compared to the 0.8 cm$^{-1}$
splitting in NH$_3$. Because of that the inversion transition is
effectively mixed with rotational transitions and the
inversion-rotational spectrum of hydronium is observed in the
submillimeter-wave region. Here we show that all hydronium
isotopologues have mixed transitions with high sensitivity to
$\mu$-variation and we also give improved estimate of the
sensitivity coefficients for the main isotopologue H$_3$O$^+$.

\section{General formalism}
\subsection{Sensitivity coefficients}
The dimensionless sensitivity coefficients to the variation
of fundamental constants can be defined as~\cite{KozLapLev10}
 \begin{equation}
   \frac{\delta \omega}{\omega} = Q_\alpha \frac{\delta \alpha}{\alpha}
   + Q_\mu \frac{\delta \mu}{\mu} + Q_g \frac{\delta g_n}{g_n}.
   \label{Eq:del_om}
 \end{equation}
Here $\alpha \approx 1/137.036$ is the fine-structure constant,
$\mu=m_e/m_p\approx 1/1836.15$, $g_n$ is the nuclear $g$-factor, and
$Q_\alpha$, $Q_\mu$, and $Q_g$ are the corresponding sensitivity
coefficients. Note that $g_n$ is not a fundamental constant, but it
weakly depends on the quark masses \cite{FT06} and has to be
considered as independent parameter whenever magnetic hyperfine
structure is involved.

It is known that tunneling and rotational transitions in molecules,
built from light elements, are mostly sensitive to $\mu$-variation:
$Q_\mu \gtrsim 1$, while $Q_\alpha \ll 1$ and $Q_g \ll 1$. In the
following we disregard coefficients $Q_\alpha$ and $Q_g$ and
concentrate on calculating the dominant sensitivity coefficient
$Q_\mu$. In the next two sections we consider purely inversion
transitions. Note, that for the asymmetric isotopologues H$_2$DO$^+$
and HD$_2$O$^+$ such transitions are not observable; they occur only
in combination with rotational transitions
$\omega_r$~\cite{KozLapLev10}. We will discuss mixed
inversion-rotational transitions in Sec.\ \ref{sec_mixed}.

\subsection{Inversion potential and reduced mass}

As explained above, we can link the variation of the inversion
transition $\delta \omega_{\rm inv}/\omega_{\rm inv}$ to the
variation of the single fundamental constant $\mu$:
 \begin{equation}
 \frac{\delta \omega_{\rm inv}}{\omega_{\rm inv}}
 = Q_{\mu,\mathrm{inv}} \frac{\delta \mu}{\mu} .
 \label{Eq:del_om1}
 \end{equation}
We can present the Hamiltonian for the inversion process as (if not
said otherwise, we use atomic units throughout the paper):
 \begin{equation}
 H = -\frac{1}{2M} \frac{\partial^2}{\partial x^2} + U(x),
 \label{Eq:H}
 \end{equation}
where $x$ is the tunneling coordinate and $M$ is the respective
reduced mass. For example, $x$ can be the distance from oxygen to
the plane of hydrogens. Following~\cite{SwaIbe62} we parameterize
the potential $U(x)$ as follows
 \begin{equation}
 U(x) = \frac{1}{2}k x^2 + \frac{1}{2}d x^4 + b\, e^{-c x^2}.
 \label{Eq:U}
 \end{equation}
Parameters $k$, $b$, $c$, and $d$ can be found by fitting vibrational
frequencies for the molecule H$_3$O$^+$. As a result we obtained $k =
0.08917$, $b = 0.04977$, $c = 1.36954$, and $d = 0.00138$.

Another way to find the parameters in \eref{Eq:U} is to fit the
inversion frequency of the ground vibrational state and the barrier
height. The latter was accurately determined within
``semi-empirical'' approach in Ref.~\cite{DonNes06} to be 652.9
cm$^{-1}$. In this case we got $k = 0.08907$, $b = 0.04967$, $c =
1.30429$, and $d =-0.00541$. To find the barrier height $\Delta U
\equiv U_{\rm max} - U_{\rm min}$ we need to solve equation
$dU(x)/dx = 0$. Obviously, $U_{\rm max} = U(0) = b = 10901$
cm$^{-1}$. As to $x_{\rm min}$ and $U_{\rm min}$, we need to solve
numerically the equation
\begin{equation}
 k + 2 d\, x_{\rm min}^2 -2 b\, c\,e^{-cx_{\rm min}^2} = 0\,,
\label{Eq:dU}
\end{equation}
which gives $x_{\rm min} = 0.5635$ and $U_{\rm min} = 10248$
cm$^{-1}$. From this we get $\Delta U \approx 653$ cm$^{-1}$ in a
perfect agreement with the semi-empirical result 652.9 cm$^{-1}$
from Ref.~\cite{DonNes06}.

Comparing two sets of parameters we see that they are very close
with the exception of the parameter $d$. We have checked that both
sets of parameters for the potential $U(x)$ lead to very close
results for the sensitivity coefficients. The numerical results
which will be discussed in the following are obtained with the
second set.

In order to find eigenvalues of the Hamiltonian (\ref{Eq:H}) we need
to know the reduced masses for different isotopologues of hydronium.
For the symmetric species H$_3$O$^+$ and D$_3$O$^+$ three protons
(or deuterons) can be treated as a single particle with the mass $m
= 3\,m_{p(d)}$. Then the reduced mass for the inversion mode can be
found within two simple models.

If we assume that the H--H distances are constant during inversion,
then the vibration involves only a change in the distance between
the plane of hydrogens and oxygen. The reduced mass $M$(H$_3$O$^+)$
can be written as
\begin{equation}
M({\rm H}_3{\rm O}^+) = \frac{3m_p M_{\rm O}}{3m_p + M_{\rm O}},
\label{Eq:M1}
\end{equation}
where $M_O$ is the mass of the oxygen nucleus.

Alternatively, we can assume that the H--O distances are constant.
Then the inversion coordinate corresponds to the angle and the
reduced mass is given by~\cite{SwaIbe62}
\begin{equation}
 M({\rm H}_3{\rm O}^+)
 = \frac{3m_p \,(M_{\rm O} + 3m_p\, {\rm sin}^2\theta)}{3m_p + M_{\rm O}}.
\label{Eq:M2}
\end{equation}
Here $\theta $ is the angle between the plane of the hydrogen atoms
and an H--O bond. Using the value of the angle, $\alpha$, between
two H--O bonds $\alpha$(HOH) = 111.3$^\circ$~\cite{SeaBunDav85} we
obtain $\theta \approx 17.6^\circ$. To find the reduced mass of the
D$_3$O$^+$ molecule we only need to change $m_p$ to $m_d\approx
2m_p$ in Eqs.~(\ref{Eq:M1}) and (\ref{Eq:M2}). Considered here
models present two limiting cases. The difference between them
constitutes only 1.7\% for H$_3$O$^+$ and 3.4\% for D$_3$O$^+$. More
accurate values for reduced mass should lie between these limits.

The motion of the asymmetric species H$_2$DO$^+$ and HD$_2$O$^+$ is
more complex and finding reduced masses is more difficult. We can
consider it as a free parameter that can be found from fitting the
theoretical inversion frequency of the ground state to the
experimental value. Note that the inversion frequencies of the
ground states are measured with high precision for all the molecules
H$_3$O$^+$, H$_2$DO$^+$, HD$_2$O$^+$, and
D$_3$O$^+$~\cite{DonNes06}.

\subsection{Sensitivity coefficient for inversion transition}
\label{sec_inv}

A numerical integration of the Schr\"{o}dinger equation with the
potential \eqref{Eq:U} allows us to find eigenvalues of the
Hamiltonian and, respectively, to calculate the inversion frequency.
Then we can find the sensitivity coefficients $Q_{\mu,\mathrm{inv}}$
by numerical differentiation taking into account that reduced mass
$M$ in atomic units scales as $\mu^{-1}$.

\begin{table}
\caption{The reduced masses $M$, the experimental and theoretical
inversion frequencies $\omega_{\rm inv}$, and the numerically
calculated sensitivity coefficients $Q_{\mu,\mathrm{inv}}$.}

\label{T:Q}

\begin{ruledtabular}
\begin{tabular}{ccccc}
&& \multicolumn{2}{c}{\underline{$\omega_{\rm inv}$ (cm$^{-1}$)}} & \\
  Molecule  & $M$ (a.u.) & Experim. & Theor. & $Q_{\mu,\mathrm{inv}}$ \\
\hline
H$_3$O$^+$  & 4639 & 55.35 & 55.79 & 1.97 \\
            & 4657 & & 55.35 & 1.98 \\
            & 4718 & & 53.94 & 1.99 \\[0.05cm]
H$_2$DO$^+$ & 5419 & 40.52 & 40.52 & 2.15 \\[0.05cm]
HD$_2$O$^+$ & 6485 & 27.03 & 27.03 & 2.36 \\[0.05cm]
D$_3$O$^+$  & 8012 & 15.36 & 15.92 & 2.65 \\
            & 8120 & & 15.36 & 2.67 \\
            & 8287 & & 14.54 & 2.70 \\
\end{tabular}
\end{ruledtabular}
\end{table}

In~\tref{T:Q} we present the results of the numerical calculations
for the ground state inversion frequencies and the sensitivity
coefficients for the isotopologues of hydronium. For the symmetric
species H$_3$O$^+$ and D$_3$O$^+$ we present results for three
values of the reduced mass. The first and the third values
correspond to Eqs.~(\ref{Eq:M1}) and (\ref{Eq:M2}), correspondingly.
The second value is found by fitting the ground state tunneling
frequency. As seen from the table the coefficients
$Q_{\mu,\mathrm{inv}}$ are almost insensitive to these small changes
of the reduced masses. Even for the heavier isotopologue D$_3$O$^+$
the differences between calculated values of $Q_{\mu,\mathrm{inv}}$
do not exceed 2\%.

Another way to find the sensitivity coefficients
$Q_{\mu,\mathrm{inv}}$ is based on the semi-classical
Wentzel-Kramers-Brillouin (WKB) approximation. A detailed
description of this approach and its application to calculations of
$Q_{\mu,\mathrm{inv}}$ for different molecules was repeatedly
discussed earlier (see,
e.g.,~\cite{SLP92,FlaKoz07,KozLapLev10,KozLev11}). For this reason
we present here only the main expressions.

Following Landau and Lifshitz~\cite{LanLif97} we can write the
inversion frequency as
\begin{equation}
\omega_{\rm inv} \approx \frac{2 E_0}{\pi}\,e^{-S},
\label{Eq:w_inv}
\end{equation}
where $S$ is the action over classically forbidden region and $E_0$
is the ground state vibrational energy calculated from the bottom of
the well $U_{\rm min}$. Harmonic approximation $2 E_0=\omega_v$
($\omega_v$ is experimentally observed vibrational frequency) is not
applicable to hydronium and $E_0$ has to be found by
solving~\eref{Eq:H} numerically. After that using the experimentally
known $\omega_{\rm inv}$ we can obtain the action $S$
from~\eref{Eq:w_inv}.

It was shown in~\cite{FlaKoz07,KozLev11} that the sensitivity
coefficient $Q_{\mu,\mathrm{inv}}$ can be expressed through $S$,
$E_0$, and $\Delta U$ as follows
\begin{equation}
 Q_{\mu,\mathrm{inv}} \approx \frac{1+S}{2} + \frac{S\, E_0}{2(\Delta
 U-E_0)}\,.
 \label{Eq:Q_WKB}
\end{equation}

The third term in~\eref{Eq:Q_WKB} was first obtained in
Ref.~\cite{FlaKoz07}. It was shown there that the numerical factor
for this term depends on the form of the potential barrier. (It is
worth mentioning that there is a typo in Eq.(12) of~\cite{FlaKoz07}:
the coefficient 1/4 should read as 1/2.) The square and triangular
potential barrier were considered. For both of them the coefficients
in front of the third term were calculated and the average was
taken. This average value is reproduced in \eref{Eq:Q_WKB}. Here we
considered a more realistic, parabolic form of the potential barrier
and found, that solution for this case exactly coincides with
\eref{Eq:Q_WKB}.

\begin{table}
\caption{The reduced masses $M$, the ground state vibrational energy
$E_0$, the action over classically forbidden region $S$, and the
sensitivity coefficients $Q_{\mu,\mathrm{inv}}$ obtained in the WKB
approximation.}

\label{T:WKB}

\begin{ruledtabular}
\begin{tabular}{ccccc}
  Molecule  & $M$ (a.u.)
                   & $E_0$ (cm$^{-1}$)
                         &  S   & $Q_{\mu,\mathrm{inv}}$ \\
\hline
H$_3$O$^+$  & 4639 & 339 & 1.36 & 1.92 \\
            & 4657 & 339 & 1.36 & 1.91 \\
            & 4718 & 337 & 1.36 & 1.90 \\[0.05cm]
H$_2$DO$^+$ & 5419 & 322 & 1.62 & 2.10 \\[0.05cm]
HD$_2$O$^+$ & 6485 & 302 & 1.96 & 2.33 \\[0.05cm]
D$_3$O$^+$  & 8012 & 279 & 2.45 & 2.64 \\
            & 8120 & 278 & 2.44 & 2.63 \\
            & 8287 & 276 & 2.44 & 2.61 \\
\end{tabular}
\end{ruledtabular}
\end{table}

Knowing the values of $S$, $E_0$, and $\Delta U$ we can calculate
the sensitivity coefficients $Q_{\mu,\mathrm{inv}}$ in the WKB
approximation. The results are presented for all molecules
in~\tref{T:WKB}. The WKB approximation is known to work well for $S
\gg 1$. As seen from~\tref{T:WKB}, for hydronium $S$ is close to
unity. For this reason we cannot anticipate high accuracy for the
sensitivity coefficients obtained in this approximation.
Nevertheless, comparing the results of Tables~\ref{T:Q} and
\ref{T:WKB}, we see a reasonable agreement between the numerical and
semi-classical values of $Q_{\mu,\mathrm{inv}}$. The largest
discrepancy does not exceed 5\%. Based on the comparison of all
approximations, considered here, we present the recommended values
of the sensitivity coefficients $Q_{\mu,\mathrm{inv}}$
in~\tref{T:Qrec}. We estimate the accuracy of our calculations to be
not worse than 5\%.
\begin{table}
\caption{The recommended values of the sensitivity coefficients
$Q_{\mu,\mathrm{inv}}$. The uncertainties are given in parentheses.}

\label{T:Qrec}

\begin{ruledtabular}
\begin{tabular}{ccccc}
    & H$_3$O$^+$ & H$_2$DO$^+$ & HD$_2$O$^+$ & D$_3$O$^+$ \\
\hline
$Q_{\mu,\mathrm{inv}}$
    &   2.0(1)   &     2.2(1)  &    2.4(1)   &    2.7(1)
\end{tabular}
\end{ruledtabular}
\end{table}

The coefficient $Q_{\mu,\mathrm{inv}}$ for H$_3$O$^+$ calculated in
this work differs by approximately 20\% from that
in~\cite{KozLev11}. In Ref.~\cite{KozLev11} this coefficient was
found in two ways: in the semi-classical approximation and from the
plot where the inversion frequency was presented as a function of
the reduced mass of hydronium isotopologues. The reduced masses were
estimated on the basis of Ref.\ \cite{DonNes06}.

We already mentioned above that the accuracy of the semi-classical
approach for H$_3$O$^+$ is not very high as the action $S\sim 1$. In
addition, the answer strongly depends on the value of $E_0$ in
\eref{Eq:Q_WKB}. Here we found $E_0$ solving an eigenvalue problem
for \eref{Eq:H}, while in Ref.~\cite{KozLev11} $E_0$ was extracted
from Ref.~\cite{DonNes06} to be approximately equal to 400
cm$^{-1}$. The second, graphical method used in~\cite{KozLev11}
strongly depends on the values of the reduced masses, which were
approximately determined from Ref.\ \cite{DonNes06}. These
approximations caused overestimation of the $Q$-factor.

Here, in order to improve the accuracy, we have done numerical
calculations within the simple one-dimensional model described
above. Optimizing parameters of the potential $U(x)$ we were able to
reproduce the experimental low-lying energy levels (the first set of
the parameters) and the transition frequency and the barrier height
(the second set of the parameters) with an accuracy better than 1\%.
However, we still estimate the accuracy of this simple model for the
$Q$-factors to be about 5\%.

\subsection{Sensitivity coefficients for mixed transitions}
\label{sec_mixed}

Sensitivity coefficients for mixed inversion-rotational transitions
in H$_3$O$^+$ were considered in \cite{KozLev11}. Here we extend
this discussion to all four isotopologues of hydronium. Rotational
motion is described by the Hamiltonian of the non-rigid asymmetric
top. We write it in the basis set $|J,K,s\rangle$ of the oblate
symmetric top ($K\equiv K_C$). Additional quantum number $s$
corresponds to the symmetric ($s=+$) and asymmetric ($s=-$)
inversion state. Rotational parameters $A^s$, $B^s$, and $C^s$
depend on the quantum number $s$. This dependence contributes to the
inversion transition $+ \leftrightarrow -$ and can be considered as
centrifugal corrections to the inversion frequency $F$.

The effective inversion-rotational Hamiltonian is diagonal in
quantum numbers $J$ and $s$ and mixes states with $K$ and $K\pm 2$.
Matrix elements diagonal in $K$ have the form:
\begin{subequations}\label{H_eff}
\begin{align}
 \label{H_eff_a}
 &H_{K,K} =-\frac{s}{2} F\\
 \label{H_eff_b}
 &+ \frac12 \left(A^s+B^s\right)
 \left[J(J+1)-K^2\right] + C^sK^2\\
 \label{H_eff_c}
 &+\Delta_J\left[J(J+1)\right]^2
 -\Delta_{KJ}J(J+1)K^2
 -\Delta_K K^4\,,
\end{align}
where the last three terms describe centrifugal corrections to rigid
rotor. Matrix elements nondiagonal in $K$ are given by:
\begin{align}
 H_{K,K+2} &= \frac14 \left(A^s-B^s\right)
 \left[J(J+1)-K(K+1)\right]^{1/2}
 \nonumber\\
 \label{H_eff_d}
 &\times\left[J(J+1)-(K+1)(K+2)\right]^{1/2}\,.
\end{align}
\end{subequations}

Effective Hamiltonian \eqref{H_eff} has ten parameters, which have
to be fitted to the experimental spectrum. For the symmetric species
we require that $A^s=B^s$ and $\Delta_{KJ}=0$, leaving seven free
parameters. Such fits were done for H$_3$O$^+$ in Ref.\
\cite{YDPP09}, for H$_2$DO$^+$ in \cite{DonNes06,FSA07,MDN10}, for
HD$_2$O$^+$ in \cite{DUD05,FuSa08}, and for D$_3$O$^+$ in
\cite{AOS99}. In the cited literature the centrifugal terms
\eqref{H_eff_c} are also assumed to depend on $s$. These
$s$-dependent corrections appear to be much smaller than all other
parameters of the effective Hamiltonian and here we reduce the
number of free parameters by neglecting them.

In order to calculate sensitivity coefficients for mixed transitions
we need to specify how parameters of the effective Hamiltonian
depend on $\mu$. The inversion transition energy is given by the
parameter $F$, whose dependence on $\mu$ was discussed in Sec.\
\ref{sec_inv}. Averaged rotational constants $A$, $B$, and $C$
depend on the respective moments of inertia and are, therefore,
proportional to $\mu$ ($A=(A^+ +A^-)/2$, etc.). Centrifugal
parameters $\Delta_i$ appear in the second order of adiabatic
perturbation theory and are quadratic in $\mu$.

It is more difficult to determine the $\mu$-dependence of the
differences $A^+ -A^-$, $B^+ -B^-$, and $C^+ -C^-$, which present
centrifugal corrections to the inversion transition energy $F$. All
of them must have the same dependence on $\mu$, so we will consider
only $W_K\equiv C^+ -C^-$.

Clearly, the rotational constant $C$ generally depends on the
inversion coordinate: $C=C_0+C_1 x^2+\dots$. This generates a
correction to the inversion potential \eqref{Eq:U}, $U(x)
\rightarrow U(x)+C_1 x^2 K^2$. To a first approximation we can
substitute $x$ with its equilibrium value $x_\mathrm{min}$.
Consequently, $U_\mathrm{min}\rightarrow U_\mathrm{min}+C_1
x_\mathrm{min}^2 K^2$. We can now use \eref{Eq:w_inv} to estimate
the respective change in inversion frequency
$\omega_\mathrm{inv}\rightarrow\omega_\mathrm{inv}+W_K K^2$. After
some algebra we arrive at:
\begin{align}
 \label{W_K}
 W_K &=\omega_\mathrm{inv}\,\frac{S} {\Delta U-E_0}\, C_1 x_\mathrm{min}^2 .
\end{align}
Differentiating this expression in respect to $\mu$ we get:
\begin{align}
 \label{dW_K}
 \frac{\delta W_K}{W_K} &=\left(Q_{\mu,\mathrm{inv}}+ \frac12\right)
 \frac{\delta\mu}{\mu}\,.
\end{align}
We conclude that centrifugal corrections to inversion frequency
scale as $\mu^{Q_{\mu,\mathrm{inv}}+ 1/2}$, which is sufficiently
close to the estimate \cite{FlaKoz07}.

\begin{table}[hbt]
\caption{Sensitivities of the low frequency mixed
inversion-rotational transitions in symmetric isotopologues of
hydronium. Molecular states are labeled with quantum numbers
$J_K^s$. Error bars for sensitivity coefficients $Q_\mu$ correspond
to the errors for inversion transitions in \tref{T:Qrec}.}
 \label{T:h3o}

\begin{ruledtabular}
\begin{tabular}{ccrrr}
 \multicolumn{2}{c}{Transition}
 &\multicolumn{2}{c}{Frequency (MHz)}
 &\multicolumn{1}{c}{$Q_\mu$}\\
 \multicolumn{1}{c}{Upper}
 &\multicolumn{1}{c}{Lower}
 &\multicolumn{1}{c}{Theory}
 &\multicolumn{1}{c}{Exper.}\\
 \hline\\[-3mm]
 \multicolumn{5}{c}{H$_3$O$^+$}\\
 $ 1_1^- $&$ 2_1^+ $&  307072 &   307192.4 &$  6.4(5) $\\
 $ 3_2^+ $&$ 2_2^- $&  365046 &   364797.4 &$ -3.5(5) $\\
 $ 3_1^+ $&$ 2_1^- $&  389160 &   388458.6 &$ -3.1(4) $\\
 $ 3_0^+ $&$ 2_0^- $&  397198 &   396272.4 &$ -3.0(4) $\\
 $ 0_0^- $&$ 1_0^+ $&  984690 &   984711.9 &$  2.7(2) $\\
 $ 4_3^+ $&$ 3_3^- $& 1031664 &  1031293.7 &$ -0.6(2) $\\
 $ 4_2^+ $&$ 3_2^- $& 1071154 &  1069826.6 &$ -0.5(2) $\\
 $ 3_2^- $&$ 3_2^+ $& 1621326 &  1621739.0 &$  2.0(1) $\\
 $ 2_1^- $&$ 2_1^+ $& 1631880 &  1632091.0 &$  2.0(1) $\\
 $ 1_1^- $&$ 1_1^+ $& 1655832 &  1655833.9 &$  2.0(1) $\\[1mm]
 \multicolumn{5}{c}{D$_3$O$^+$}\\
 $ 3_1^- $&$ 3_1^+ $&  450608 &   450709.7 &$  2.7(1) $\\
 $ 3_2^- $&$ 3_2^+ $&  454910 &   454940.3 &$  2.7(1) $\\
 $ 2_1^- $&$ 2_1^+ $&  456194 &   456211.4 &$  2.7(1) $\\
 $ 1_1^- $&$ 1_1^+ $&  459918 &   459917.7 &$  2.7(1) $\\
 $ 2_2^- $&$ 2_2^+ $&  460496 &   460493.1 &$  2.7(1) $\\
 $ 3_3^- $&$ 3_3^+ $&  462080 &   462074.6 &$  2.7(1) $\\
 $ 0_0^- $&$ 1_0^+ $&  120117 &            &$  7.5(4) $\\
 $ 2_1^+ $&$ 1_1^- $&  220408 &            &$ -2.5(2) $\\
 $ 2_0^+ $&$ 1_0^- $&  221782 &            &$ -2.5(2) $\\
\end{tabular}
\end{ruledtabular}
\end{table}

\begin{table}[hbt]
\caption{Sensitivities of the low frequency mixed
inversion-rotational transitions in asymmetric isotopologues of
hydronium. Molecular states are labeled with quantum numbers
$J_{K_A,K_C}^s$.}
 \label{T:h2do}

\begin{ruledtabular}
\begin{tabular}{ccrrr}
 \multicolumn{2}{c}{Transition}
 &\multicolumn{2}{c}{Frequency (MHz)}
 &\multicolumn{1}{c}{$Q_\mu$}\\
 \multicolumn{1}{c}{Upper}
 &\multicolumn{1}{c}{Lower}
 &\multicolumn{1}{c}{Theory}
 &\multicolumn{1}{c}{Exper.}\\
 \hline\\[-3mm]
 \multicolumn{5}{c}{H$_2$DO$^+$}\\
 $ 3_{1,2}^+ $&$ 2_{0,2}^- $&  210994 &   211108.8 &$ -5.9(6) $\\
 $ 1_{0,1}^- $&$ 2_{1,1}^+ $&  250920 &   250914.1 &$  6.8(5) $\\
 $ 3_{2,1}^+ $&$ 2_{1,1}^- $&  312737 &   312831.8 &$ -3.6(4) $\\
 $ 3_{2,2}^+ $&$ 2_{1,2}^- $&  412156 &   412130.2 &$ -2.5(3) $\\
 $ 3_{3,0}^+ $&$ 2_{2,0}^- $&  632799 &   632901.7 &$ -1.2(2) $\\
 $ 3_{3,1}^+ $&$ 2_{2,1}^- $&  649742 &   649653.4 &$ -1.2(2) $\\
 $ 0_{0,0}^- $&$ 1_{1,0}^+ $&  673229 &   673257.0 &$  3.2(2) $\\
 $ 4_{2,2}^+ $&$ 3_{1,2}^- $&  715955 &   715827.9 &$ -1.0(2) $\\
 $ 4_{1,3}^+ $&$ 3_{0,3}^- $&  716961 &   716959.4 &$ -1.0(2) $\\
 $ 1_{1,1}^- $&$ 2_{2,1}^+ $&    6633 &            &$ 219(18) $\\
 $ 1_{1,0}^- $&$ 2_{2,0}^+ $&   51108 &            &$   29(2) $\\[1mm]
 \multicolumn{5}{c}{HD$_2$O$^+$}\\
 $ 0_{0,0}^- $&$ 1_{1,0}^+ $&  380753 &   380538.0 &$  4.0(2) $\\
 $ 3_{2,2}^+ $&$ 2_{1,2}^- $&  474792 &   474541.1 &$ -1.4(2) $\\
 $ 5_{2,3}^- $&$ 5_{3,3}^+ $&  525419 &   525451.5 &$  3.1(2) $\\
 $ 3_{3,0}^+ $&$ 2_{2,0}^- $&  610712 &   610573.1 &$ -0.8(1) $\\
 $ 3_{3,1}^+ $&$ 2_{2,1}^- $&  626907 &   627069.8 &$ -0.8(2) $\\
 $ 6_{2,4}^- $&$ 6_{3,4}^+ $&  632532 &   632776.9 &$  1.9(1) $\\
 $ 3_{1,2}^- $&$ 3_{2,2}^+ $&  634305 &   633793.2 &$  2.8(1) $\\
 $ 4_{1,3}^- $&$ 4_{2,3}^+ $&  707925 &   707552.6 &$  2.6(1) $\\
 $ 1_{0,1}^- $&$ 1_{1,1}^+ $&  728086 &   728420.2 &$  2.6(1) $\\
 $ 1_{0,1}^- $&$ 2_{1,1}^+ $&   35047 &            &$   33(2) $\\
 $ 2_{2,0}^+ $&$ 1_{1,0}^- $&   93734 &            &$  -11(1) $\\
 $ 2_{2,1}^+ $&$ 1_{1,1}^- $&  129633 &            &$ -7.7(6) $\\
\end{tabular}
\end{ruledtabular}
\end{table}

Since we determined how the parameters of the effective Hamiltonian depended
on $\mu$, we can find the sensitivity of the mixed transitions by numerical
differentiation of the eigenvalues of the effective Hamiltonian. Results of
these calculations are listed in Tables \ref{T:h3o} and \ref{T:h2do}. Note
that if we neglect all centrifugal corrections, we can write the approximate
expressions for the frequency and the sensitivity of the mixed
inversion-rotational transition \cite{KozLapLev10}:
\begin{subequations}\label{appox_mix}
\begin{align}
 \label{w_mix}
 \omega_\mathrm{mix}
 &= \omega_r \pm \omega_\mathrm{inv}\,,\\
 \label{Q_mix}
 Q_{\mu,\mathrm{mix}}
 &= \frac{\omega_r}{\omega_\mathrm{mix}}
 \pm Q_{\mu,\mathrm{inv}}
 \frac{\omega_\mathrm{inv}}{\omega_\mathrm{mix}}\,.
\end{align}
\end{subequations}
For small values of the quantum numbers $J$ and $K$ considered here the
centrifugal corrections are small and approximation \eqref{appox_mix} is
rather good.

\section{Discussion}\label{dusc}

Our final results for sensitivity coefficients for
inversion-rotational transitions between low-lying states of all
hydronium isotopologues are listed in Tables~\ref{T:h3o} and
\ref{T:h2do}. There we also give calculated frequencies and
experimental frequencies, where known (see Refs.\
\cite{VTM88,AOS99,FSA07,FuSa08}). The errors for the $Q$-factors
correspond to the uncertainties in \tref{T:Qrec}. For low-frequency
mixed transitions the errors are increased due to cancelations
between rotational and tunneling contributions in \eref{Q_mix}. For
the experimentally observed transitions we use experimental
frequencies to calculate $Q$-factors. For unknown transitions we
have to use predicted values. The accuracy of our model with limited
number of centrifugal corrections decrease with increasing
rotational energy. For given $J$ the latter is smaller for heaviest
isotopologue. Therefore, the accuracy of the model is highest for
D$_3$O$^+$ ($\sim 100$ MHz) and lowest for H$_3$O$^+$, where the
maximum error is almost 1 GHz. However, the errors in $Q$-factors
from the inaccuracy in frequency do not exceed 2\% and can be
neglected compared to the error associated with
$Q_{\mu,\mathrm{inv}}$.

It is clear from \eref{appox_mix} that mixed transitions can have
high sensitivities only if $|\omega_\mathrm{mix}|
\ll\omega_\mathrm{inv}$ [note, that \eqref{w_mix} allows negative
frequencies $\omega_\mathrm{mix}$]. Thus, we are primarily
interested in transitions from the microwave range and low-frequency
submillimeter transitions. One can see from
Tables~\ref{T:h3o} and \ref{T:h2do}, that all isotopologues have
such transitions with both positive and negative sensitivities,
whose absolute values are significantly larger than unity.

The lowest frequency 6.6 GHz corresponds to the transition
$1^-_{1,1} \rightarrow 2^+_{2,1}$ in H$_2$DO$^+$. This frequency is
184 times smaller than respective inversion frequency. Consequently,
the sensitivity coefficient for this transition is close to two
hundred. Other frequencies are at least an order of magnitude larger
and respective sensitivities are much smaller.

As we pointed out above, our present value of $Q_{\mu,\mathrm{inv}}$
for H$_3$O$^+$ is 20\% smaller than that in Ref.\ \cite{KozLev11}.
Consequently, the absolute values of our $Q$-factors for mixed
transitions are also significantly smaller. Still, the difference
$\Delta Q$ between sensitivities of the 307 GHz line on the one hand
and 365 GHz and 396 GHz lines on the other hand is close to 10, or 3
times larger, than $\Delta Q$ for ammonia method \cite{FlaKoz07}.
These three lines have been observed from the interstellar medium
(see \cite{vdT06} and references therein). That makes hydronium very
promising candidate for the $\mu$-variation search.

Submillimeter spectra of H$_3$O$^+$ are usually observed from warm
and dense star forming regions, where hydronium is one of the most
abundant molecule \cite{BBvD10}. Up to now deuterated hydronium was
not observed from the interstellar medium. However, deuterated
isotopologues of ammonia were observed several times \cite{LGR08}.
Thus, it is possible that hydronium isotopologues can be found in
the future. As we have showed here, this can give additional
possibilities to study $\mu$-variation.

\section{Acknowledgments}
We are grateful to Sergei Levshakov for bringing the isotopologues of
hydronium to our attention. This work was supported by The Deutsche
Forschungsgemeinschaft SFB 676, Teilprojekt C4 and by RFBR grant No.\
11-02-00943.


\end{document}